\documentclass[manuscript,screen]{acmart}
\AtBeginDocument{%
  }

\usepackage[linesnumbered,ruled,vlined,noend]{algorithm2e} 

\usepackage[most]{tcolorbox}
\newtcolorbox{mybox}{
  colback=gray!10,     
  colframe=black!75,   
  boxrule=0.5pt,       
  arc=2pt,             
  outer arc=2pt,
  left=6pt,            
  right=6pt,           
  top=6pt,             
  bottom=6pt,          
  boxsep=0pt,          
}

\usepackage{tabularx}   
\usepackage{adjustbox}
\usepackage{booktabs}
\usepackage{threeparttable}
\usepackage{array}
\usepackage{enumitem}
\newcolumntype{Y}{>{\centering\arraybackslash}X} 
\usepackage{makecell}
\usepackage{bm}
\usepackage{amsmath}

\newcommand{\mytodored}[1]{\textcolor{red}{\ding{46}~{\sf}~#1}}

\usepackage{pgffor}
\usepackage{pifont}

\newif\ifshowcomments
\showcommentstrue
\ifshowcomments
\newcommand{\wu}[1]{\mytodored{[wu: #1]}}
\newcommand{\hhx}[1]{\mytodored{[hhx: #1]}}
\newcommand{\other}[1]{\mytodored{[xiangyang: #1]}}
\else
\newcommand{\wu}[1]{}
\newcommand{\hhx}[1]{}
\newcommand{\other}[1]{}
\fi

\begin{document}

\title{LDMDroid: Leveraging LLMs for Detecting Data Manipulation Errors in Android Apps}

\author{Xiangyang Xiao}
\affiliation{
  \institution{Xiamen University}
  \city{Xiamen}
  \state{Fujian}
  \country{China}
}
\orcid{0009-0006-0352-6739}
\email{xiangyangxiao@stu.xmu.edu.cn}

\author{Huaxun Huang}
\affiliation{
  \institution{Xiamen University}
  \city{Xiamen}
  \state{Fujian}
  \country{China}
}
\orcid{0000-0002-1778-3721}
\authornote{Corresponding author.}
\email{huanghuaxun@xmu.edu.cn}

\author{Rongxin Wu}
\affiliation{
  \institution{Xiamen University}
  \city{Xiamen}
  \state{Fujian}
  \country{China}
}
\orcid{0000-0002-4648-3795}
\email{wurongxin@xmu.edu.cn}

\begin{abstract}
Android apps rely heavily on Data Manipulation Functionalities (DMFs) for handling app-specific data through CRUDS operations, making their correctness vital for reliability. However, detecting Data Manipulation Errors (DMEs) is challenging due to their dependence on specific UI interaction sequences and manifestation as logic bugs. Existing automated UI testing tools face two primary challenges: insufficient UI path coverage for adequate DMF triggering and reliance on manually written test scripts. To address these issues, we propose an automated approach using Large Language Models (LLMs) for DME detection. We developed LDMDroid, an automated UI testing framework for Android apps. LDMDroid enhances DMF triggering success by guiding LLMs through a state-aware process for generating UI event sequences. It also uses visual features to identify changes in data states, improving DME verification accuracy. We evaluated LDMDroid on 24 real-world Android apps, demonstrating improved DMF triggering success rates compared to baselines. LDMDroid discovered 17 unique bugs, with 14 confirmed by developers and 11 fixed. The tool is publicly available at \href{https://github.com/runnnnnner200/LDMDroid}{https://github.com/runnnnnner200/LDMDroid}.
\end{abstract}




\maketitle

\section{Introduction} \label{sec:intro}

Android apps have become an indispensable part of modern life,
enabling a diverse array of daily activities,
from communication and entertainment to finance and productivity~\cite{AndroidStats2025}.
Among the various features of Android apps, data manipulation functionalities (DMFs) are prevalent~\cite{marianiAugustoExploitingPopular2018,sunPropertyBasedFuzzingFinding2023}.
DMFs handle app-specific data following CRUDS operations (a.k.a. create, read, update, delete, and search), such as adding notes, retrieving records, and removing files.
Ensuring the correctness of DMFs is critical, as they typically serve as the foundation for an app's core functionalities. Failure to do so can lead to data manipulation errors (DMEs).

\begin{figure}[t]
    \centering
    \includegraphics[width=0.75\textwidth]{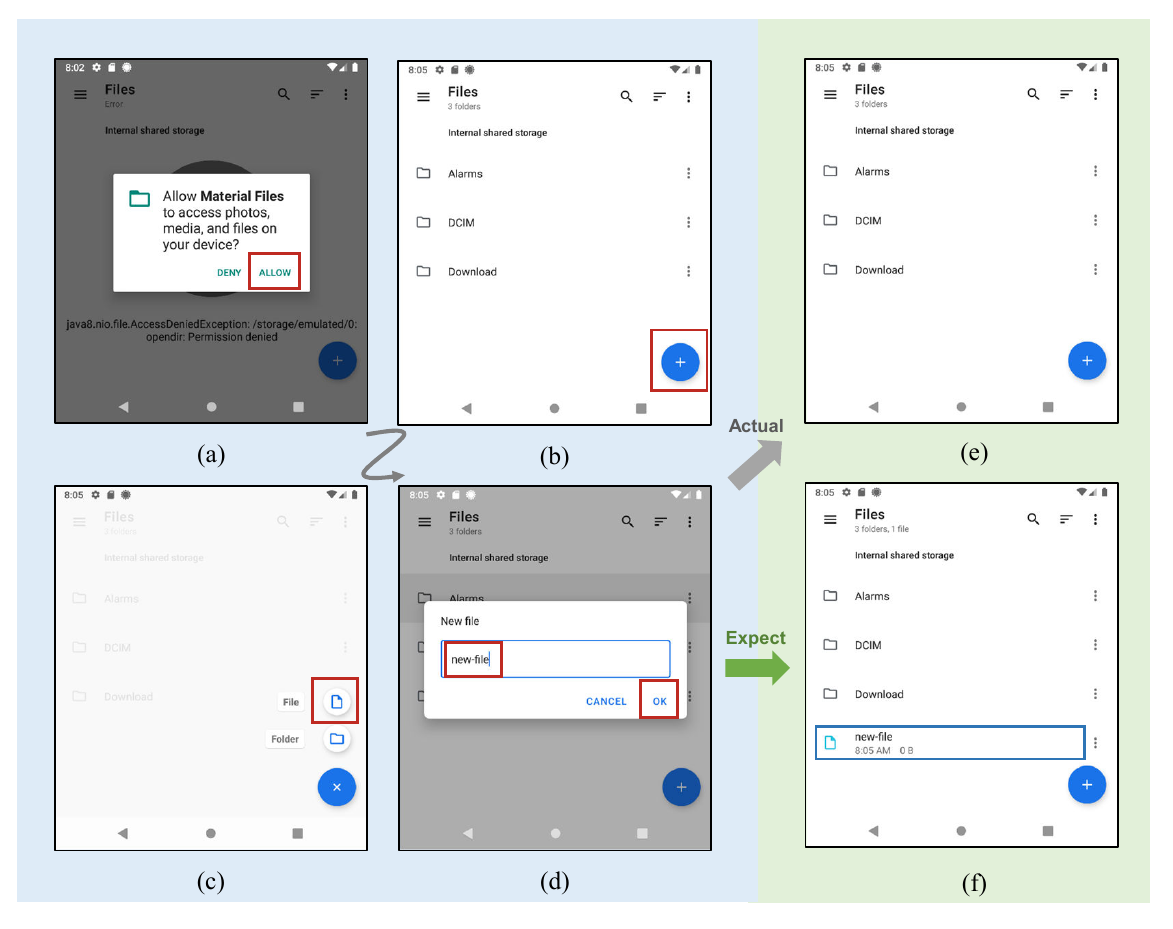}
    \caption{A DME in \textit{Material Files} (v1.7.4) related to ``Create File'' DMF. The small red boxes indicate UI events.
    The expected outcome (f) shows the newly created file appearing in the file list,
    while the actual outcome (e) reveals its absence, demonstrating the DME.}
    \Description{Screenshots illustrating a DME in the Material Files app during the file creation process. (a) shows the user granting storage permissions, (b) to (d) depict creating a new file named ``new-file''. The expected result in (f) shows the file appearing in the list, while (e) shows the file missing from the list, indicating the failure and the DME.}
    \label{fig:real-example}
\end{figure}

Fig.~\ref{fig:real-example} illustrates a real-world DME that occurs during the file creation process in \textit{Material Files}~\cite{MaterialFiles}.
\textit{Material Files} is a file management app for Android devices,
where the ``create file'' functionality is considered a DMF,
and the file list represents the data under manipulation (DUM).
During the execution of this DMF, a DME occurred that negatively impacted the user experience, as shown in Fig.~\ref{fig:real-example}.
Specifically, the user first grants storage permissions when prompted (see Fig.~\ref{fig:real-example}(a)),
then creates a new file named ``new-file'' under the default directory (see Fig.~\ref{fig:real-example}(b)$\sim$(d)).
As expected, the file ``new-file'' should be visible in the file list (see blue box in Fig.~\ref{fig:real-example}(f)).
However, the file did not appear in the file list (Fig.~\ref{fig:real-example}(e)), indicating that the file creation operation had failed, causing a DME that disrupted the expected behavior of the file creation DMF.

For app developers, detecting DMEs in Android apps poses the following challenges:
(1) There is no unified paradigm that can comprehensively cover all possible DMFs across apps. Therefore, triggering such issues depends on app-specific DMFs with different semantics.
For example, in Fig.~\ref{fig:real-example}, the process of creating a new file constitutes a multi-step DMF with app-specific semantics:
tapping the ``Create'' icon, selecting the ``File'' option, entering a filename such as ``new-file'',
and confirming the action by tapping the ``OK'' button.
(2) These bugs are typically non-crashing bugs and are often detected by developers using manually written test scripts. However, detecting such bugs is both time-consuming and labor-intensive. Therefore, there is an urgent need for an automated approach to detect these DMEs in Android apps. 

Currently, existing approaches exhibit the following limitations in detecting DMEs.
First, general UI-based testing approaches for Android apps~\cite{androidMonkey2025,liDroidBotLightweightUIguided2017, suGuidedStochasticModelbased2017a, guPracticalGUITesting2019, wangComboDroidGeneratingHighquality2020a, liHumanoidDeepLearningbased2020, panReinforcementLearningBased2020a, lvFastbot2ReusableAutomated2023} often struggle to effectively trigger DMFs that require an understanding of the apps' semantics in CRUDS operations.
Second, although several approaches have been proposed to trigger and detect DMEs in Android apps~\cite{sunPropertyBasedFuzzingFinding2023, xiongGeneralPracticalPropertybased2024}, they require considerable manual effort to construct test scripts that can both trigger and validate DMFs, which significantly limits their practicality in real-world apps.
Notably, Gu et al. proposed DMSDroid~\cite{guDMSDroid2024}, which attempts to automatically discover DMFs. 
However, this approach primarily relies on hardcoded rules and lacks a deep understanding of app-specific semantics, and therefore fails to discover the UI event sequences triggering DMFs in different Android apps. For example, in case of DMF for Create operations, DMSDroid encodes a hardcoded rule as follows: clicking a button containing the word ``add'', editing the content in an input field, and finally clicking a button containing the word ``submit''. However, such a rule clearly fails to uncover all necessary UI events depicted in Fig.~\ref{fig:real-example} (a.k.a., selecting the type of file to be created as shown in Fig.~\ref{fig:real-example}(c)).

In recent years, large language models (LLMs) have demonstrated capabilities in understanding the semantics of text, code, and images, which can aid in understanding the semantics of apps' UI ~\cite{zhaoSurveyLargeLanguage2025}.
So far, research efforts have explored leveraging LLMs for automated Android UI testing~\cite{
wenAutoDroidLLMpoweredTask2024, ranGuardianRuntimeFramework2024, songVisionTaskerMobileTask2024,wangMobileAgentESelfEvolvingMobile2025, liuTestingLimitsUnusual2024, liuMakeLLMTesting2024}. Motivated by this, we choose to investigate how to leverage LLMs for the detection of DMEs.

However, despite providing domain-specific knowledge, such as the general steps for triggering DMFs and the definition of DMEs as described by Sun et al.~\cite{sunPropertyBasedFuzzingFinding2023}, to existing LLM-based approaches~\cite{ranGuardianRuntimeFramework2024, songVisionTaskerMobileTask2024, liuMakeLLMTesting2024}, these approaches still face the following challenges:
\begin{itemize}[leftmargin=*, noitemsep, topsep=4pt]
\item
\textcolor{black}{
\textbf{DMF Correctness Verification.}
Verifying the correctness of a DMF is challenging because constructing test oracle of such logical bugs requires precisely capturing the key semantic information from the large semantic space involved in the data manipulation process. For example, as shown in Fig.~\ref{fig:real-example}(f), the logical oracle must not only detect the newly added list item but also verify that it corresponds to the created file. Although the LLM has certain capabilities in understanding app semantics~\cite{wenDroidBotGPTGPTpoweredUI2024, wangEnablingConversationalInteraction2023}, it still struggles to identify key semantic changes and generate reliable test oracles for logical bugs.
}
\item 
\textcolor{black}{
\textbf{UI Action Planning.}
While planning UI actions for triggering DMFs, contextual information can be important to ensure the completeness of the task. However, contextual information can vary in different forms, such as the UI hierarchy or the text shown on UI components, making it challenging to identify which aspects are important. Existing approaches~\cite{ranGuardianRuntimeFramework2024, liuMakeLLMTesting2024} typically record only action logs as the contextual information. For example, in Fig.~\ref{fig:real-example}(b), the click on the ``Add'' icon button is recorded as \textit{``click the ImageButton with text sd\_main\_fab''}. Such logs omit the semantic role of the action, making it difficult for the LLM to correctly understand the action history and plan subsequent actions.}
\end{itemize}

\textcolor{black}{
In view of the above challenges, we propose LDMDroid, an LLM-based approach to facilitate the automated exploration of DMFs and detection of DMEs in Android apps.
To tackle the first challenge, we model the test oracle as \textit{comparing the state changes of data lists before and after triggering DMFs} (e.g., the ``file list'' in Fig.~\ref{fig:real-example}(b), which shows file entries as a list). Such an observation is derived from our empirical study of numerous real-world DMEs.
Our insight into the second challenge is that \textit{UI changes provide important semantic information about the outcomes of historical UI actions, enabling LLMs to better understand the current state when exploring app UIs}. For example, in Fig.~\ref{fig:real-example}(b), clicking the ``Add'' button brings up a creation menu. Providing this UI change as contextual information helps LDMDroid recognize that the app has entered the file creation state and subsequently trigger the DMF.}

We implemented and evaluated LDMDroid on 24 open-source Android apps.
Experimental results demonstrate that our approach significantly improves the success rate of triggering DMFs (see Section~\ref{sec:RQ1-RQ2}). By modeling data list state changes and tracking UI changes, the effectiveness of LDMDroid is improved by 55.9\% and 87.6\%, respectively.
In addition, LDMDroid uncovered 17 previously-unknown bugs across 10 apps.
All of these bugs were reported to the respective developers, with 14 confirmed and 11 already fixed.
Notably, a substantial portion of these bugs (13 out of 17) could not be detected by the state-of-the-art automated approaches (see Section~\ref{sec:RQ3}).

In summary, this paper makes the following contributions:
\begin{itemize}[leftmargin=*, noitemsep, topsep=4pt]
    \item To the best of our knowledge, we are the first to investigate how LLMs can be leveraged to validate the functional correctness of DMFs.
    \item \textcolor{black}{
    To improve DMF identification, we summarize common DMF patterns and design a state-aware mechanism that enhances LLMs in planning UI actions by leveraging feedback from UI changes.
    }
    \item We implemented LDMDroid and evaluated it on real-world Android apps, and the results demonstrate its effectiveness and practical utility in detecting DMEs.
\end{itemize}

\section{Data Manipulation}

In an Android app, a UI page can be represented by a tree-structured view hierarchy \( H \),
where each node in \( H \) corresponds to a UI widget \( w \), which contains a set of attributes (e.g., \texttt{android:text}) to configure the visual appearance of the UI widget.
A UI event is defined as \( e = (t, w, o) \),
where \( t \) denotes the event type (e.g., \texttt{Click}, \texttt{InputText}, \texttt{Back}),
\( w \) refers to the target widget that the event acts upon (which may be empty, such as in a \texttt{Back} event),
and \( o \) is an optional data field (e.g., the text input in an \texttt{InputText} event).

At runtime, Android apps visualize data through widgets on UI pages.
A sequence of UI events is required to trigger corresponding DMFs,
which allow app users to manipulate the data displayed on the screen. 
Once a DMF is triggered, the DUM and its corresponding UI widget typically show relevant changes.
For example, in the case illustrated in Fig.~\ref{fig:real-example},
executing the ``Create File'' DMF should result in the addition of a new file to the file list widget,
with the filename matching the text input by the user during the DMF execution process.

Following the formal definition of Sun et al.~\cite{sunPropertyBasedFuzzingFinding2023}, we abstract a DMF as a Hoare Triple: \(\{Pre\}\; E \; \{Post\}\)
where \( Pre \) denotes the precondition,
\( E \) is the sequence of UI events triggering the functionality, denoted as \( E = [e_1, \dots, e_i, \dots, e_n] \),
with \( e_i \) representing a UI event, and \( Post \) represents the postcondition.
If the current app state satisfies \( Pre \),
we can send the event sequence \( E \) to trigger the DMF,
and then verify \( Post \) to check the correctness of the DMF. 
We categorize DMFs into five types,
with the definitions of their semantics, Pre, and Post conditions outlined in Table~\ref{tab:dmf-types}.
Moreover, we consider the DUM as a container that is responsible for displaying a set of data \( D = [d_1, \dots, d_i, \dots, d_n] \). 
For example, in the ``create file'' DMF, \( D \) corresponds to the file list. 
DME happens when the app does not behave as specified in \( Post \). For example, in Fig.~\ref{fig:real-example}(e), the target file does not appear in the file list,
indicating that the file creation functionality failed to execute properly. 

\begin{table}[t]
	\centering
	\small
	\caption{Preconditions, Semantics, and Postconditions of DMFs}
	\label{tab:dmf-types}
	\begin{threeparttable}
		\begin{tabularx}{\textwidth}{r|l|l|l}
			\toprule
			\textbf{Type} & \textbf{Semantics of} $\boldsymbol{E}$ & $\boldsymbol{Pre}$ & $\boldsymbol{Post}$ \\
			\midrule
			Create & Create a new data item in DUM & $D \in L_0 \land e_{t}.w \in L_0$ & $D' \in L_n \land d_{t} \in D' \land |D'| = |D| + 1$ \\
			Update & Modify an existing data item in DUM & $D \in L_0 \land |D| > 0 \land e_{t}.w \in L_0$ & $D' \in L_n \land d_{t} \in D \land |D'| = |D|$ \\
			Delete & Delete a data item from DUM & $D \in L_0 \land |D| > 0 \land e_{t}.w \in L_0$ & $D' \in L_n \land d_{t} \notin D \land |D'| = |D| - 1$ \\
			Read & View details of a data item in DUM & $D \in L_0 \land |D| > 0 \land e_{t}.w \in L_0$ & The semantics of $d_{t}$ matches the view in $L_n$ \\
			Search & Search for a data item in DUM & $D \in L_0 \land |D| > 0 \land e_{t}.w \in L_0$ & The semantics of $d_{t}$ matches the view in $L_n$ \\
			\bottomrule
		\end{tabularx}
		\begin{tablenotes}
		\footnotesize
		\item $e_{t}$ denotes the first event in $E$ with a non-empty $w$ field, $d_{t}$ is the data item manipulated by the UI actions, \textcolor{black}{
        $D$ and $D'$ represent the same data container before and after the data manipulation.
        }
		\end{tablenotes} 
	\end{threeparttable}	
\end{table}
\section{Motivation}
\label{sec:motivation}

Sun et al.~\cite{sunPropertyBasedFuzzingFinding2023} proposed PBFDroid, which is the first tool in exploring the detection of DMEs in Android apps.
However, the bug detection process of PBFDroid is semi-automated and still relies on manual effort in providing valid UI event sequences of DMFs and test oracle.
To the best of our knowledge, DMSDroid, proposed by Gu et al.~\cite{guDMSDroid2024},
is currently the state-of-the-art automated method for detecting DMEs.
DMSDroid works by building a UI transition graph (UTG) of the target app, and then tries to match  potential UI event sequences triggering DMF on UTG.
However, DMSDroid has the following limitations:
(1) DMSDroid applies hardcoded rules to match DMF execution paths and lacks a deep understanding of app-specific UI semantics, often failing to discover potential DMFs (see Section~\ref{sec:intro});
(2) DMSDroid is limited to detecting bugs that cause app crashes, making it unable to identify non-crashing functional bugs that result from DMF execution.

In view of the potential capabilities of LLMs in understanding the semantics of UI pages, we explore leveraging LLMs to automatically plan the UI event sequences required to trigger DMFs and to verify the logical correctness of the DMF execution.
Existing work has been proposed to leverage LLMs for testing Android apps~\cite{wenAutoDroidLLMpoweredTask2024, ranGuardianRuntimeFramework2024, songVisionTaskerMobileTask2024}.
Among them, Guardian~\cite{ranGuardianRuntimeFramework2024} is the state-of-the-art approach that evaluates each generated UI action by checking whether it can be helpful to trigger the specified functionalities in Android apps.
We adapted Guardian for the detection of DMEs in two ways. First, we converted the semantics and validation logic of each DMF type in Table~\ref{tab:dmf-types} into natural-language descriptions and embedded them into its prompts, enabling Guardian to leverage DMF knowledge during interactions to guide action planning and validate the correctness of a DMF.
Second, we enhanced Guardian with UI screenshots as visual modality inputs to provide additional visual context during interaction.
Despite these improvements, Guardian still faces challenges in reliably triggering and validating certain DMFs.

\begin{itemize}[leftmargin=*, noitemsep, topsep=4pt]
\item \textbf{Challenge 1: DMF Correctness Verification.}
\textcolor{black}{
Verifying the correctness of a DMF relies on constructing a logical oracle over its execution results, i.e., determining whether the performed data manipulation conforms to the expected logical semantics. However, the apps consist of diverse semantic information, such as different UI components and actions. Therefore, accurately capturing and verifying the key semantics that are relevant to DMF correctness becomes challenging. Although the LLM has certain capabilities in understanding app semantics, our experimental results in Section~\ref{sec:RQ3} show that relying solely on the LLM still makes it difficult to correctly identify the critical semantic changes during DMF execution and generate reliable logical oracles. For example, as shown in Fig.~\ref{fig:real-example}(f), after executing a file creation DMF, a new entry appears in the file list. In this case, a correct logical oracle should not only detect the presence of a newly added list item in the UI, but also further confirm that the entry indeed corresponds to the file created by the UI action.
}

\item \textbf{Challenge 2: UI Action Planning.}
\textcolor{black}{
LDMDroid requires the LLM to make reasonable decisions about the next action. However, the contextual information involved in this process can be varied, such as UI structures or interaction histories. Therefore, selecting the key information as the contextual information is challenging. Existing approaches~\cite{ranGuardianRuntimeFramework2024, liuMakeLLMTesting2024} typically record only historical action logs, which leads to the loss of important semantic information. For example, in Fig.~\ref{fig:real-example}(b), the click on the ``Add'' icon button is recorded as \textit{``click the ImageButton with text sd\_main\_fab''}. With only such action descriptions, the LLM may fail to recognize the functional semantics triggered by the click (e.g., initiating a creation workflow), which can lead to decisions that do not align with the current task state. On the other hand, directly preserving the complete exploration history (e.g., sequences of UI screenshots) introduces substantial redundant information for LLMs.
}
\end{itemize}

LDMDroid tackles the above challenges as follows:
\begin{itemize}[leftmargin=*, noitemsep, topsep=4pt]
\item \textcolor{black}{
To address Challenge 1, our key insight is that \textit{the correctness of a DMF is mainly related to the state changes of the DUM}, which typically presents multiple structurally similar data items in a list. This insight is grounded in our empirical analysis of 81 real-world DMFs documented by Sun et al.~\cite{sunPropertyBasedFuzzingFinding2023}. Inspired by this observation, we abstract a DUM as a structured data container and transform DMF semantic correctness verification into a problem of comparing state transitions of this container. This abstraction provides more focused contextual information for LLM to generate oracles. For instance, in Fig. \ref{fig:real-example}(b), the DUM corresponds to the ``file list''. LDMDroid identifies this file list and then extracts its state representations before and after the file creation. By comparing these two state representations, LDMDroid can determine whether the newly created file has been correctly inserted into the list, thereby verifying the correctness of the DMF.
}
\item \textcolor{black}{To address Challenge 2, our insight is that \textit{the UI changes caused by UI actions provide key semantic information for supporting subsequent planning}. These UI changes reflect the historical progress, thereby enabling the LLM to perceive the current task state and plan appropriate subsequent UI actions. To achieve this goal, LDMDroid extracts the UI changes resulting from each UI action, such as widget updates, content changes, layout adjustments, or screen transitions. LDMDroid further leverages these UI changes to track task progress and provide state-aware historical context for subsequent UI action planning. For example, the action in Fig.~\ref{fig:real-example}(b) leads to the UI changes shown in (c), which is recorded as: \textit{``after clicking the file button, a dialog for file creation appears.''}. With the above information, LDMDroid can recognize that the current task has entered the \textit{Enter the creation information} stage and generate the UI action for entering the file name.}

\end{itemize}
\section{Approach}

\begin{figure}[t]
    \centering
    \includegraphics[width=0.8\textwidth]{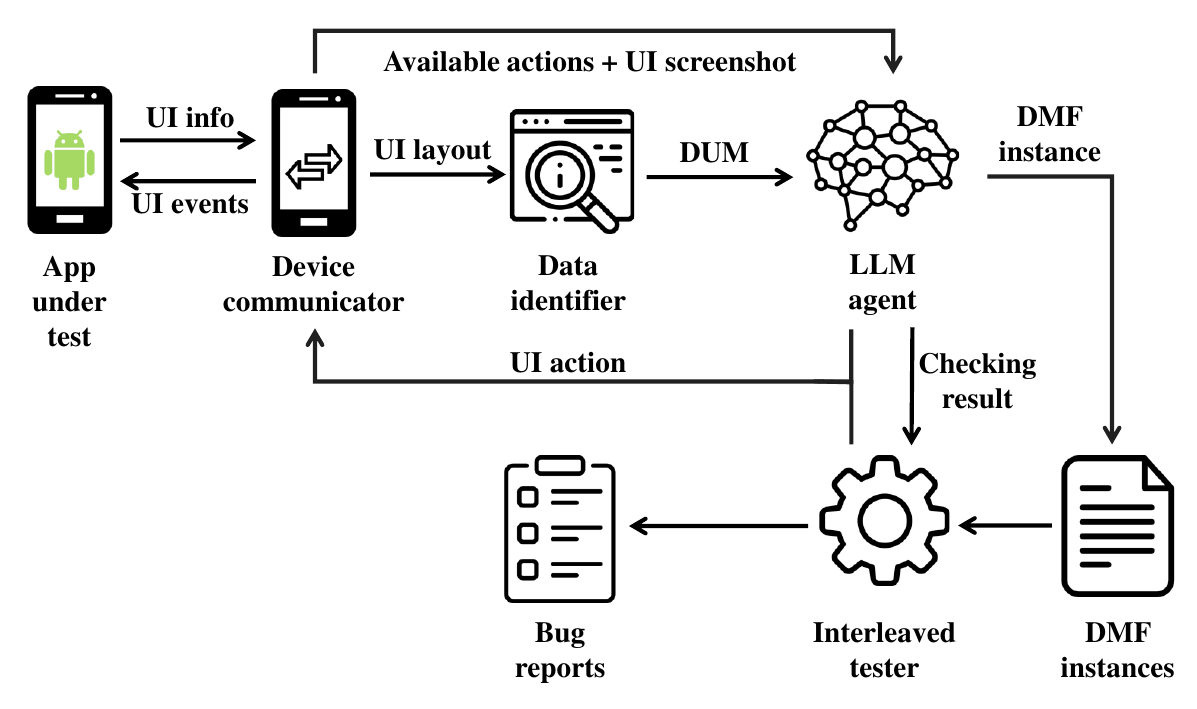}
    \caption{The Overview of LDMDroid.}
	\Description{The figure shows LDMDroid's workflow: it first uses random UI events to find DUMs, then employs the LLM to generate UI event sequences that trigger DMFs. Finally, it combines these events with others to detect potential DMEs and produce bug reports.}
    \label{fig:workflow-overview}
\end{figure}

In this paper, we introduce LDMDroid, a method based on LLMs for detecting potential DMEs in Android apps. Fig.~\ref{fig:workflow-overview} provides an overview of LDMDroid.
Initially, LDMDroid takes the app under test (AUT) as input and explores it under the guidance of an LLM to identify DUMs.
Next, LDMDroid leverages the LLM to collect DMF instances by generating UI event sequences that are capable of triggering the DMFs associated with the identified DUMs.
Finally, LDMDroid interleaves the triggering of DMFs with other UI events in a randomized manner to detect potential DMEs and generate bug reports.

\subsection{Identifying DUMs} \label{sec:recognition-of-dum}

\begin{center}
	\begin{algorithm}[t]
		\caption{Identify DUMs in UI Page}
		\label{alg:identify-dum}
		\KwIn{Tree-structured UI hierarchy $\mathcal{T}$ of a UI page}
		\KwOut{Set of clusters $\mathcal{C} = \{C_1, \dots, C_n\}$ containing DUMs}
		\SetAlgoLined
		
		Initialize an empty set of clusters $\mathcal{C}$\;
		\ForEach{subtree $w$ in $\mathcal{T}$}{
			\ForEach{existing cluster $C_i$ in $\mathcal{C}$}{
				\ForEach{widget $w_j$ in $C_i$}{
					\If{$sim_{structure}(w, w_j) \leq \text{threshold}_{structure}$ \textbf{and} $sim_{align}(w, w_j) \leq \text{threshold}_{align}$}{
						Add $w$ to cluster $C_i$\;
						\textbf{break}\;
					}
				}
			}
			\If{not added to any cluster}{
				Create cluster $C_{new} = \{w\}$ and add it to $\mathcal{C}$\;
			}
		}
		
		\ForEach{cluster $C_i$ in $\mathcal{C}$}{
			\If{all widgets in $C_i$ display string constants}{
				Remove $C_i$ from $\mathcal{C}$\;
			}
		}
		\Return $\mathcal{C}$\;
	\end{algorithm}
\end{center}

To help LLMs accurately identify DUMs, we conducted an empirical analysis of 81 DMF instances collected from 20 apps by Sun et al.~\cite{sunPropertyBasedFuzzingFinding2023}.
We found that all DUMs are contained in data containers and displayed on the UI page with a similar structure.
Based on this observation, LDMDroid identifies data containers composed of structurally similar items as DUMs. For example, the file list widget shown in Fig.~\ref{fig:real-example} is a DUM, characterized by displaying multiple structurally similar items in a list format.
Such DUM identification serves as the foundation for verifying DMF correctness by enabling reliable comparison of UI states on data containers, as described in Section~\ref{sec-test-oracle}.

Algorithm~\ref{alg:identify-dum} describes how LDMDroid identifies DUMs within a UI page. Starting with the tree-structured UI hierarchy \( \mathcal{T} \), it initializes clusters \( \mathcal{C} \). Widgets are grouped into clusters based on their structural similarity \( sim_{structure} \) and positional alignment \( sim_{align} \).
\textcolor{black}{
The structural similarity \( sim_{structure} \) quantifies the differences between the subtrees of two widget $w_i$ and $w_j$ as follows:
}
\textcolor{black}{
\begin{equation}
sim_{structure}(w_i,w_j)=
\frac{Dist(w_i,w_j)}{\max(|N_i|,|N_j|)}
\label{eq:sim-structure}
\end{equation}
$Dist(w_i,w_j)$ denotes the tree edit distance, where nodes are considered identical if they belong to the same widget class, and $|N_i|,|N_j|$ denote the numbers of nodes in the two subtrees.
}
\textcolor{black}{
The positional alignment \( sim_{align} \) measures the spatial alignment between two widgets, considering both their positions and size differences. It is calculated as follows:
\begin{equation}
sim_{align}(w_i,w_j)=
\min \left(
|x_i-x_j| + \lambda |h_i-h_j|,\;
|y_i-y_j| + \lambda |w_i-w_j|
\right)
\label{eq:sim-align}
\end{equation}
}
\textcolor{black}{
$(x_i,y_i)$ denotes the center coordinates of widget $w_i$, and $(w_i,h_i)$ denotes its bounding box size.
$\lambda$ is a weighting factor for the size difference and is set to $0.5$ in our implementation.
For both metrics, smaller values indicate higher similarity.
The thresholds are empirically set to $threshold_{structure}=0.2$ and $threshold_{align}=100$.
If a widget does not fit into existing clusters, a new cluster is created (Lines 1-9). 
}

After constructing preliminary clusters, LDMDroid further filters out those clusters that consist entirely of widgets displaying string constants. The filtering process is necessary because such widgets, like the ``cancel'' and ``ok'' button in Fig.~\ref{fig:real-example}(d), typically serve as labels of apps' functionalities and do not contain dynamic content. The filtering process involves decompiling the app to identify string constants and exclude clusters composed of these string constants only (Lines 10--13).

LDMDroid leverages an LLM to explore the app and initialize user data in a human-like manner. Specifically, the LLM is prompted to behave as a normal user who navigates the app to create data. At each step, the system provides the LLM with the current UI screenshot and a structured description of available UI widgets along with the interaction types supported by each widget (e.g., \texttt{Click} or \texttt{InputText}).
Based on this information, LDMDroid leverages the LLM to decide which UI widget to interact with and what interaction to perform, and the system executes the corresponding UI action.
After each interaction, the system applies the DUM identification algorithm to the resulting UI page to detect potential DUMs.
Once a DUM is identified, the system records the DUM and saves the corresponding app snapshot for subsequent DMF instance collection (see Section~\ref{sec:collect-dmf}).

\begin{algorithm}[t]
\caption{Collect DMF Instances}
\label{alg:collect-dmfs}
\KwIn{$DUM$ in a UI page of $AUT$}
\KwOut{$DMFs$}

\SetAlgoLined
\SetKw{KwBreak}{break}
\SetKw{KwAnd}{and}
\SetKwFunction{FInit}{initializeGeneralSteps}
\SetKwFunction{FPlan}{planAndExecuteNextAction}
\SetKwFunction{FUIChange}{extractUIChanges}
\SetKwFunction{FCheckProgress}{checkTaskProgress}
\SetKwFunction{FValidate}{validateDMF}
\SetKwFunction{FRestoreAppState}{restoreAppSnapshot}
\SetKwFunction{FPop}{pop}
\SetKwFunction{FDiscoverSiblingGoals}{discoverSiblingGoalsOnce}
\SetKwProg{Fn}{Function}{}

\ForEach{$operation \in \{\text{Create}, \text{Update}, \text{Delete}, \text{Read}, \text{Search}\}$}{
    \textcolor{black}{
    $goals \gets \{initialGoal\}$\;
    }
    \While{\textcolor{black}{$goals \neq \emptyset$}}{
        \FRestoreAppState{}\;
        \textcolor{black}{
        $goal \gets$ \FPop{$goals$}\;
        }
    	\FInit{$operation$};~\tcp{Section \ref{sec-DMF-general-steps}}
    	\For{$step \gets 1$ \KwTo MAX\_STEP\_LENGTH}{
            \textcolor{black}{
            $action \gets$ \FPlan{$goal$};~\tcp{Section \ref{sec-DMF-plan-actions}}
            }
    		\FUIChange{$action$};~\tcp{Section \ref{sec-DMF-extract-ui-changes}}
    		$progress \gets $\FCheckProgress{};~\tcp{Section \ref{sec-DMF-track-progress}}
    		\If{$progress ==$ COMPLETE}{
                \If{\FValidate{}~\tcp{Section \ref{sec-test-oracle}}}{
                    Add $dmf$ to $DMFs$\;
                    \textcolor{black}{
                    $goals \gets goals \cup$ \FDiscoverSiblingGoals{$dmf$};~\tcp{Section \ref{sec-discover-sibling-goal}}
                    }
                }
                \KwBreak\;
    		}
    	}
    }
}
\Return $DMFs$\;
\end{algorithm}

\subsection{Collecting DMF Instances}  \label{sec:collect-dmf}

LDMDroid further collects DMF instances after DUM identification. Algorithm~\ref{alg:collect-dmfs} illustrates this process.
\textcolor{black}{
For each DMF operation type, LDMDroid initializes a high-level objective (Line 2), denoted as \textit{initialGoal}, representing the operation's default manipulation semantics. For example, under the Create type, \textit{initialGoal} specifies performing a Create operation on the identified DUM.
}
At the beginning of each attempt, LDMDroid restores the app to the previously recorded snapshot, resetting it to the state in which the identified DUM is ready for DMF collection (Line 4). 
Given the general steps for triggering DMFs of the current type (Line 6), LDMDroid iteratively monitors the task progress. In each iteration, it provides the LLM with the next general step and the history of previous UI actions as contextual information to plan the subsequent UI action. LDMDroid then evaluates the UI changes caused by each UI action and provides this feedback to the LLM to guide further UI action planning (Lines 8–10). This attempt continues until LDMDroid determines that the DMF has been successfully triggered (Lines 11-16), or the UI event sequence reaches the maximum length \textit{MAX\_SETP\_LENGTH} (Line 7).
If the DMF is successfully triggered, LDMDroid validates its correctness (Line 8), and only DMF instances that pass this validation are preserved for subsequent testing.
\textcolor{black}{
Additionally, after the first successful collection of a DMF instance for a given operation type, LDMDroid analyzes the associated action history to identify other sibling DMF goals (Line 15). Newly discovered goals are incorporated into subsequent attempts to facilitate targeted collection of related DMF instances (Line 5).
}

\begin{table}[t]
	\centering
	\caption{General Steps for Triggering DMFs}
	\label{tab:dmf-workflows}
	\begin{tabular}{>{\centering\arraybackslash}p{2.5cm}|p{5.5cm}}
		\toprule
		\textbf{Type} & \textbf{General steps} \\
		\midrule
		Create & \makecell[l]{
			1. Open the create page \\
			2. Choose the creation type \\
			3. Enter the creation information \\
			4. Submit and save \\
			5. Return
		} \\
		\midrule
		Update & \makecell[l]{
			1. Select the data to edit \\
			2. Open the edit page \\
			3. Enter the modification information \\
			4. Submit and save \\
			5. Return
		} \\
		\midrule
		Delete & \makecell[l]{
			1. Select the data to delete \\
			2. Delete the data \\
			3. Confirm the deletion \\
			4. Return
		} \\
		\midrule
		Read & \makecell[l]{
			1. Select the data to read \\
			2. Open the detail page \\
			3. Read the detail
		} \\
		\midrule
		Search & \makecell[l]{
			1. Open the search page \\
			2. Enter the search keyword \\
			3. Read the search result
		} \\
		\bottomrule
	\end{tabular}
\end{table}

\subsubsection{General Steps for Triggering DMFs} \label{sec-DMF-general-steps}

To design the general steps, three authors of this paper independently analyzed the 81 DMF instances collected by Sun et al.~\cite{sunPropertyBasedFuzzingFinding2023} and the five DMF extraction rules proposed by Gu et al.~\cite{guDMSDroid2024}. After individual analyses, all the authors discussed and reached a consensus on the general steps required to trigger different types of DMFs. By comparing these general steps with the collected instances, we found that they cover 88.9\% of the actual steps. Based on this, we derived unified general steps for five DMF types, as shown in Table~\ref{tab:dmf-workflows}.
\textcolor{black}{
A comparison with the DMFs manually identified during dataset construction further indicates that the actual triggering procedures are largely consistent with these general steps. On average, 77.6\% of the actual steps required for each DMF are covered by the corresponding general steps.
}
It is worth noting that LTGDroid does not rely on a strict match to the general steps. Instead, it leverages the LLM’s generalization capability to flexibly handle potential deviations between the general and actual triggering procedures.

\begin{table}[t]
	\centering
	\caption{Prompt Template of LDMDroid}
	\label{tab:prompt-template}
	\begin{threeparttable}
		\begin{tabular}{l|l|l}
			\toprule
			\textbf{Usage} & \textbf{Context} & \textbf{Instruction} \\
			\midrule
			Plan the next action & \makecell[l]{
				1. Available UI actions \\
				2. Action history \\
				3. Current progress in general steps
			} & Next action to trigger DMF? \\
			\midrule
			Extract UI changes & \makecell[l]{
				1. Last UI action \\
				2. DUM \\
				3. Screenshots before and after action
			} & UI changes caused by the last action? \\
			\midrule
			Track task progress & \makecell[l]{
				1. Action history \\
				2. General steps
			} & Next step of the task? All done? \\
			\midrule
			Test oracle for logical DMEs & \makecell[l]{
				1. DUM before manipulation \\
				2. DUM after manipulation \\
				3. Target data of manipulation \\
				\textcolor{black}{4. User inputs} \\
				5. Manipulation result screenshot \\
				6. Definition of logical DMEs
			} & Any logic DME? Why? \\
			\midrule
			\textcolor{black}{Discover sibling DMFs} & \makecell[l]{
				\textcolor{black}{1. Action history}
			} & \textcolor{black}{Any other sibling DMFs?} \\
			\bottomrule
		\end{tabular}
		\begin{tablenotes}
			\footnotesize
			\item The complete prompt template is available in the online document~\cite{PromptTemplateLDMDroid}.
		\end{tablenotes}
	\end{threeparttable}
\end{table}

\subsubsection{Planning UI Actions} \label{sec-DMF-plan-actions}

LDMDroid then continues to plan UI actions based on the identified general steps for triggering DMFs. The prompt design is shown in Table~\ref{tab:prompt-template}. Specifically, LDMDroid takes the available UI actions, the action history, and the current progress in general steps as inputs and outputs the next UI action to trigger DMFs. Following the approach of Ran et al.~\cite{ranGuardianRuntimeFramework2024}, LDMDroid considers generating the following five types of UI actions: \texttt{Click}, \texttt{LongClick}, \texttt{InputText}, \texttt{Scroll}, and \texttt{Back}. It enumerates all UI widgets where these actions can be applied by traversing the view hierarchy of the UI page.
To avoid redundant UI actions, LDMDroid filters out actions that have already appeared in the action history when generating the current set of available UI actions. This strategy ensures that the system does not repeatedly execute the same UI action within a single attempt to trigger a DMF, preventing ineffective exploration loops. Although this strategy may theoretically miss DMFs that require repeated actions, we did not observe such cases in our evaluated dataset. Moreover, LDMDroid can be easily adapted by setting a maximum execution count for individual actions to support such scenarios if needed.

\subsubsection{Extracting UI Changes} \label{sec-DMF-extract-ui-changes}

As we mentioned in Section \ref{sec:motivation}, existing LLM-based approaches cannot determine whether the quality of the generated UI actions meets the requirements for triggering DMFs, leading to hallucination problems and resulting in UI actions that fail to trigger DMFs effectively. To address this issue, LDMDroid considers providing feedback on the UI changes before and after the execution of UI actions to enhance the quality of the generated UI actions. As shown in Table \ref{tab:prompt-template}, we provide the LLM with the last UI action, the DUM, and the screenshots before and after the action. The LLM then summarizes the resulting UI changes caused by the action. These summarized UI changes are incorporated into the action history, serving as contextual information for planning subsequent UI actions.

\subsubsection{Tracking the Task Progress} \label{sec-DMF-track-progress}

Although we defined general steps for various types of DMFs in Section~\ref{sec-DMF-general-steps}, relying solely on such general steps is insufficient to effectively guide the LLM in planning subsequent actions.
This is mainly because, in practice, LLMs often struggle to align the planned UI actions with the general steps, making it unclear what the next step should be or whether the process has already been completed, and thus leading to deviations of triggering DMFs.
In view of this, we enable LDMDroid to infer the current progress within the general steps using the UI changes generated in the previous step, thereby guiding the planning of subsequent UI actions.
Table \ref{tab:prompt-template} outlines the prompts used by LDMDroid. Specifically, LDMDroid leverages the action history, which includes both previous actions and their resulting UI changes, along with the general steps,
to prompt the LLM to infer the current progress relative to the general steps.
The task progress in LDMDroid advances flexibly rather than strictly step by step.
Intermediate steps can be skipped, and the process terminates once the LLM infers that the task for triggering the target DMF has been completed.
Such inferred task progress is then used as contextual information for planning subsequent UI actions.

\subsubsection{Validating the Correctness of DMFs} \label{sec-test-oracle}

\begin{table}[t]
    \centering
    \caption{Definitions of Logical DMEs as Prompts}
    \label{tab:logic-errors}
    \begin{tabular}{>{\centering\arraybackslash}p{1.5cm}|p{12cm}}
        \toprule
        \textbf{Type} & \textbf{Definition} \\
        \midrule
        Create & \makecell[l]{
			A logical error occurs if the target data was not correctly added to the data container. \\
			Focus on whether the target data appears in the data container after the operation.} \\
        \midrule
        Update & \makecell[l]{
			A logical error occurs if the target data was not correctly modified as expected. \\
			Focus on whether the target data has been updated with the correct values.} \\
        \midrule
        Delete & \makecell[l]{
			A logical error occurs if the target data was not correctly removed. \\
			Focus on whether the target data is still present or correctly removed in the data container.} \\
        \midrule
        Read & \makecell[l]{
			A logical error occurs if the target data was not correctly fetched or displayed as expected. \\
			Focus on whether the target data's details are correctly displayed.} \\
        \midrule
        Search & \makecell[l]{
			A logical error occurs if the target data does not appear in the search results as expected. \\
			Focus on whether the target data appears in the results.} \\
        \bottomrule
    \end{tabular}
\end{table}

To validate the correctness of a DMF, we design an LLM-based test oracle that examines the states of the target DUMs before and after triggering the DMF.
This design is based on the intuition that \textit{different types of DMFs should introduce observable UI state changes on their associated DUMs}.
Following the logical properties defined by Sun et al.~\cite{sunPropertyBasedFuzzingFinding2023}, we derived natural-language logical DME definitions for each DMF type, as shown in Table~\ref{tab:logic-errors}.
Because each DMF instance is linked to a specific DUM, LDMDroid can reliably capture the corresponding DUM states from the app's view hierarchy.
As shown in Table~\ref{tab:prompt-template}, the DUM states, \textcolor{black}{the user inputs}, together with the derived DME definition, are then used to construct prompts for the LLM-based test oracle.
To ensure the rigor of our approach and reduce randomness in LLM outputs, we set the temperature to 0 and execute each test case three times. LDMDroid generates a bug report when it identifies a DME in two or more of its outputs.

\subsubsection{\textcolor{black}{Discover Sibling DMFs}\label{sec-discover-sibling-goal}}

\textcolor{black}{
Each DMF type may include multiple concrete instances. For example, the Create type can involve both file creation and folder creation within a file manager app. In practice, during the collection of a DMF instance, additional potential sibling DMFs of the same type can often be identified.
Therefore, we provide the LLM with the action history of the first successfully collected DMF instance. The history records the executed actions and the associated UI changes summarized during the collection process, enabling the LLM to generate specific goal descriptions for other potential sibling DMFs (see Table~\ref{tab:prompt-template}). These generated goals serve to guide the targeted collection of sibling DMFs.
}

\subsection{Testing with DMF Instances}

LDMDroid leverages the collected DMF instances to explore the AUT and detect potential DMEs. To achieve this, we adopt the approach proposed by Sun et al.~\cite{sunPropertyBasedFuzzingFinding2023}, which interleaves the triggering of these DMF instances with randomly generated UI events.
Each triggered DMF instance is validated using the test oracle introduced in Section~\ref{sec-test-oracle}. If a DMF fails this validation, LDMDroid reports a logical DME.
In addition to logical DMEs, LDMDroid also monitors app logs throughout the exploration process to detect crash DMEs.

\section{Evaluation} \label{sec:evaluation}

We implemented LDMDroid by leveraging the prompt template as shown in Table \ref{tab:prompt-template}, and our evaluation aims to answer the following research questions:

\begin{itemize}[leftmargin=*]
\item \textbf{RQ1 (Effectiveness \& Efficiency):} How effective and efficient is LDMDroid in finding the correct UI event sequences that trigger DMFs compared to state-of-the-art techniques?
\item \textbf{RQ2 (Ablation Study):} How do UI changes summarization and progress tracking contribute to the effectiveness and efficiency of LDMDroid in finding correct UI event sequences that trigger DMFs?
\item \textbf{RQ3 (Usefulness):} How effective is LDMDroid at automatically finding DMEs in real-world Android apps compared to state-of-the-art techniques?
\item \textcolor{black}{
\textbf{RQ4 (Practicality):} How does LDMDroid compare with PBFDroid in terms of required manual effort and bug detection effectiveness?
}
\end{itemize}

Note that our goal is not to improve overall code coverage.
Instead, we aim to cover specific paths related to DMFs in order to uncover DMEs.
Therefore, this study does not evaluate code coverage metrics.

\subsection{Evaluation Setup}

\noindent \textbf{App Subjects.}
We selected open-source Android apps as the evaluation subjects for LDMDroid.
Specifically, we first crawled all apps from F-Droid, one of the largest open-source Android app repository~\cite{FDroidFreeOpen},
and filtered the ones hosted on GitHub so that we could submit issue reports. This process yielded 2,552 candidate apps.
We further filtered the apps to retain only those that are actively maintained (i.e., updated within a year) and relatively popular (i.e., with over 50 GitHub stars), resulting in 757 candidates.
Finally, to avoid potential overfitting, we excluded apps that were already included in the PBFDroid~\cite{sunPropertyBasedFuzzingFinding2023} and DMSDroid~\cite{guDMSDroid2024} datasets, leading to a set of 735 candidate apps.

We then randomly sampled 50 apps as an initial subset.
To identify the ground-truth DMFs in these apps, we recruited 12 software engineering graduate students, following prior work~\cite{sunPropertyBasedFuzzingFinding2023, zhaoReCDroidAutomaticallyReproducing2019}. All participants had at least one year of experience working with Android apps and had no prior connection with the authors. Prior study has shown that graduate students can represent professional testers in the software engineering experiments~\cite{salmanAreStudentsRepresentatives2015}. Each app was assigned to six participants, who manually explored the entire app to identify its overall behavior and the number of DMFs it contained. Any disagreements were resolved through discussion until consensus was reached.
During this process, we found 7 apps whose login or registration flows could not be bypassed via scripting and 19 apps that did not contain any DMFs and were therefore irrelevant to our goal of triggering DMEs. We excluded all such apps and ultimately obtained an evaluation dataset consisting of 24 apps, as shown in Table~\ref{tab:app-subjects}.
The selected apps are diverse with GitHub stars ranging from 56 (\textit{Rank My Favs}~\cite{RankMyFavs}) to 6.8k (\textit{Material Files}~\cite{MaterialFiles}),
with an average of 1,318.

\begin{table}[t]
    \centering
    \caption{Evaluated Apps (K=1,000)}
    \label{tab:app-subjects}
    \begin{tabular}{>{\centering\arraybackslash}p{1cm}|p{4.7cm}|>{\centering\arraybackslash}p{2cm}|>{\centering\arraybackslash}p{2.5cm}}
        \toprule
        \textbf{ID} & \textbf{App Name} & \textbf{Stars} & \textbf{Version} \\
        \midrule
        1  & Another Notes~\cite{AnotherNotes} & 378   & v1.5.4 \\
        2 & CycleStreets~\cite{CycleStreets} & 218   & v3.12.0 \\
        3 & CPU Info~\cite{CPUInfo} & 904   & v6.3.0 \\
        4  & Easy Notes~\cite{EasyNotes} & 678   & v1.4 \\
        5  & Fridgey~\cite{Fridgey} & 139   & v2.2.1 \\
        6 & Home Medkit~\cite{HomeMedkit} & 80    & v1.7.9 \\
        7  & LinkHub~\cite{LinkHub} & 214   & v1.6.1 \\
        8  & Material Files~\cite{MaterialFiles} & 6.8K  & v1.7.4 \\
        9  & Material Notes~\cite{MaterialNotes} & 217   & v1.12.1 \\
        10 & NoNonsense Notes~\cite{NoNonsenseNotes} & 395   & v7.2.0 \\
        11 & Notally~\cite{Notally} & 1.9K  & v6.1 \\
        12 & NotallyX~\cite{NotallyX} & 278   & v7.3.1 \\
        13 & OsmAnd~\cite{OsmAnd} & 5.4K  & v5.1.7 \\
        14  & PFA Todo List~\cite{PFATodoList} & 103   & v3.1.0 \\
        15 & Photo Editor~\cite{PhotoEditor} & 4.4K  & v3.0.2 \\
        16 & Play NotePad~\cite{PlayNotePad} & 99    & v1.3.7 \\
        17  & Print Notes~\cite{PrintNotes} & 65    & v0.9.14 \\
        18 & Quillpad~\cite{Quillpad} & 928   & v1.4.25 \\
        19 & Rank My Favs~\cite{RankMyFavs} & 56    & v0.6.11 \\
        20 & Read You~\cite{ReadYou} & 5.9K  & v0.12.1 \\
        21 & Recurring Expense Tracker~\cite{RecurringExpenseTracker} & 184   & v0.16.0 \\
        22 & Table Habit~\cite{TableHabit} & 639   & v1.16.7 \\
        23 & Tasky~\cite{Tasky} & 198   & v3.0.1 \\
        24 & To Don't~\cite{ToDont} & 160   & v4.0.0 \\
        \bottomrule
    \end{tabular}
\end{table}

\noindent \textbf{Baselines.} We compared LDMDroid with the following baselines:
\begin{itemize}[leftmargin=*]
\item \textbf{Guardian~\cite{ranGuardianRuntimeFramework2024}:}
an LLM-based automated framework that executes natural-language task descriptions through reflective reasoning. In our study, we adapt Guardian with necessary extensions to support DMF triggering and logical correctness verification for DME detection.
\item \textbf{DMSDroid~\cite{guDMSDroid2024}:}
an automated approach that extracts DMFs using hardcoded rules and employs Q-learning to explore apps for crash-related DMEs.
\item \textbf{Fastbot2~\cite{lvFastbot2ReusableAutomated2023}:}
an automated approach that utilizes reinforcement learning to guide test execution using learned event-to-activity transition knowledge from prior test runs for efficient crash detection.
\textcolor{black}{
\item \textbf{Genie~\cite{Genie}:}
an automated approach that leverages the independence among sibling views, aiming to detect non-crashing functional bugs.
}
\textcolor{black}{
\item \textbf{Odin~\cite{Odin}:}
an automated approach that identifies abnormal app behaviors through state-based differential analysis, designed for detecting non-crashing functional bugs.
}
\textcolor{black}{
\item \textbf{PBFDroid~\cite{sunPropertyBasedFuzzingFinding2023}:}
a semi-automated approach that leverages manually defined DMF properties to randomly interleave DMFs with other events for detecting DMEs.
}
\end{itemize}

Following the process illustrated in Section \ref{sec:motivation}, we adapted Guardian with task descriptions instructing it to trigger the target type of DMF (e.g., performing a Create operation). We further added a verification module that leverages DUM UI screenshots to assess the logical correctness of each DMF. To ensure fairness, we aligned Guardian's prompts with LDMDroid by incorporating the same natural language knowledge, including DMF definitions, logical DME definitions, and general steps for triggering DMFs.

\noindent \textbf{Runtime Environment.}
All the experiments were deployed and executed on Android emulators (generic\_x86\_64, Android 9.0, 2GB RAM, 4-core CPU) running on a 64-bit Windows 11 machine.
We configured LDMDroid and the baseline approaches using Zhipu AI's GLM-4V-Plus multimodal model~\cite{ZhiPu}.
As LDMDroid is not designed specifically for any particular LLM, we expect that it can be effectively applied to a broader range of LLMs.
\textcolor{black}{
To assess the sensitivity of LDMDroid's effectiveness to model selection, we additionally evaluate LDMDroid using different multimodal models (see Table~\ref{tab:llm-comparison}).
}

\noindent \textbf{Evaluation Setup of RQ1 and RQ2.}
We compared LDMDroid with Guardian, DMSDroid and two modified baselines of LDMDroid for an ablation study: {LDMDroid w/o UI Changes}, which removes the UI change summarization module of LDMDroid; {LDMDroid w/o Tracking Progress}, which removes the progress tracking module of LDMDroid.
Fastbot2, Genei and Odin were excluded from the comparison as they are not designed to trigger DMFs.
We conducted experiments using these tools on 24 selected apps. The evaluation aimed to identify valid UI event sequences that successfully trigger DMFs. For each of the five types of DMF, we used each tool to generate 10 UI event sequences, with each sequence having a maximum length of 10 UI actions. To minimize the impact of randomness in the LLM outputs, we repeated the experiment three times.
However, DMSDroid cannot be configured to generate a fixed number of UI event sequences that trigger specific DMFs. Therefore, we executed DMSDroid three times for each app, with each run lasting 90 minutes, in order to collect all UI event sequences that DMSDroid considers as triggering DMFs.
Two authors independently verified whether the generated UI event sequences correctly triggered the corresponding DMFs without causing side effects, with any disagreements resolved by a third author.
To assess the diversity of functionality triggered, we counted the number of distinct DMFs successfully triggered by each tool ($\#\mathrm{DMF}{\mathrm{tool}}$) and compared it against the ground-truth DMF count obtained during dataset construction ($\#\mathrm{DMF}{\mathrm{gt}}$).
Finally, each tool was evaluated along three dimensions: 
(1) \textbf{Coverage}, defined as 
$\mathrm{DMF}_{\mathrm{cov}} = \#\mathrm{DMF}_{\mathrm{tool}} / \#\mathrm{DMF}_{\mathrm{gt}}$, 
where $\#\mathrm{DMF}_{\mathrm{tool}}$ is the number of distinct DMFs successfully triggered by the tool; 
(2) \textbf{Effectiveness}, measured by the success rate 
$SR = \text{(\# of successful DMF-triggering sequences)} / \text{(\# of all generated sequences)}$; and 
(3) \textbf{Efficiency}, quantified by the average number of tokens, their corresponding monetary cost, and the time required to successfully trigger one DMF.

\noindent \textbf{Evaluation Setup of RQ3.}
To answer RQ3, we evaluated the DME detection capabilities of the baselines on the Android emulator. Specifically, we ran each approach on 24 selected app subjects for 90 minutes, repeated this process three times, and aggregated all bug reports across the three runs to obtain the final results.
We classified a bug report as a true positive (TP) only if it could be manually reproduced. To reduce human bias in validating the detected DMEs, two authors independently analyzed each bug report, and any disagreements were resolved by a third author.
To quantify the precision of LLM-based approaches, we further computed the overall true positive rate, defined as $\mathrm{TPR} = \text{(\# of true positive reports)} / \text{(\# of all reports)}$.
In addition, we introduced an LDMDroid w/o DUM variant, which prevents our approach from leveraging DUM's view hierarchy information to validate DMFs, to evaluate the contribution of the DUM identification module.

\noindent \textbf{Evaluation Setup of RQ4.}
\textcolor{black}{
To evaluate the practicality of LDMDroid, we compared it with the semi-automated approach PBFDroid from two perspectives: required manual effort and bug detection effectiveness. We randomly assigned 24 apps to 12 recruited participants, with each participant evaluating 10 apps, ensuring that every app was assessed by five different participants. Before the study, all participants received training on (1) filtering false-positive reports generated by LDMDroid and (2) writing scripts to define DMF properties required by PBFDroid.
For LDMDroid, participants manually filtered the false-positive reports generated during RQ3. For PBFDroid, participants manually defined DMF properties corresponding to all 137 identified DMFs across the 24 apps. PBFDroid was then equipped with these properties and executed under the same configuration as RQ3 (90 minutes per run, repeated three times) to obtain its bug detection results.
Finally, we compared the two approaches in terms of average manual effort measured by time spent per app and the number of detected bugs.
}

\subsection{Results for RQ1 and RQ2} \label{sec:RQ1-RQ2}

\begin{table}[t]
    \centering
    \caption{DMF Triggering Performance Comparison (K=1,000)}
    \label{tab:dmf-success-rates}
    \begin{threeparttable}
        \begin{tabular}{l|c|c|c|c|c}
            \toprule
            \textbf{Approaches} &  \textbf{$\bm{\mathrm{DMF}_{\mathrm{cov}}}$ (\%)}&  \textbf{SR (\%)} & \textbf{Tokens (K)} & \textbf{Cost (\$)} & \textbf{Time (s)}\\
            \midrule
            Guardian & 54.2 [43.8-64.5] & 29.9 [27.8-31.9] & 49.1 & 0.034 & 293.3 \\
            DMSDroid & 11.6 [0.2-22.9] & 5.9 [0.9-10.8] & - & - & - \\
            \textbf{LDMDroid} & \textbf{75.7 [62.7-88.7]} & \textbf{62.5 [57.0-68.0]} & \textbf{43.5} & \textbf{0.026} & \textbf{153.4} \\
            LDMDroid w/o UI Changes & 64.4 [51.8-76.9] & 33.3 [29.4-37.3] & 27.4 & 0.017 & 151.3 \\
            LDMDroid w/o Tracking Progress & 54.7 [44.2-65.1] & 36.3 [33.9-38.8] & 37.8 & 0.023 & 126.8 \\
            \bottomrule
        \end{tabular}
        \begin{tablenotes}
			\footnotesize
            \item The ``$\mathrm{DMF}_{\mathrm{cov}}$'' and ``SR'' columns report the per-app average together with their 95\% confidence intervals.
		\end{tablenotes}
    \end{threeparttable}
\end{table}

Table~\ref{tab:dmf-success-rates} compares the results of LDMDroid, Guardian and DMSDroid in triggering DMFs.
Based on the manual assessment of the 24 selected apps during dataset construction, we identified a total of 137 ground-truth DMFs (\#DMF$_{\mathrm{gt}}$).
In terms of effectiveness, LDMDroid achieved an average per-app DMF coverage of 75.7\% and success rate of 62.5\%.
In contrast, Guardian achieved an average per-app DMF coverage of 54.2\% and success rate of 29.9\%, and it did not discover any DMFs beyond those already identified by LDMDroid.
Unlike LDMDroid, Guardian lacks progress tracking and provides no clear indication of the next general step, making it more difficult to trigger potential DMFs within the apps. Additionally, Guardian does not have an effective method for evaluating the effectiveness of the UI actions it generates. This results in final UI event sequences that fail to successfully trigger DMFs, thus lowering both its DMF coverage and success rate.
DMSDroid achieved an average per-app DMF coverage of 11.6\% and success rate of 5.9\%, and all the DMFs it identified were also detected by LDMDroid.
The major reason is that DMSDroid applies hardcoded rules to check whether the generated UI event sequence can match the characteristics of a DMF. Such hardcoded rules lack a deep understanding of app-specific UI semantics, causing DMSDroid to miss potential DMFs.

In terms of efficiency, LDMDroid outperforms Guardian. As shown in Table~\ref{tab:dmf-success-rates}, LDMDroid requires only 43.5K tokens, \$0.026, and 153.4 seconds on average to successfully trigger one DMF, whereas Guardian requires 49.1K tokens, \$0.034, and 293.3 seconds. The primary reason for Guardian’s lower efficiency is that the UI event sequences it generates contain a large number of meaningless or redundant actions, which not only increase the token consumption and monetary cost but also significantly prolong the execution time. Since DMSDroid does not rely on the LLM, it is not included in the efficiency comparison.

We also observed failure cases that arise from the following two aspects.
First, we identified four apps in which LDMDroid failed to correctly recognize DUMs. Three of them (e.g., \textit{CycleStreets}~\cite{CycleStreets}, \textit{OsmAnd}~\cite{OsmAnd}, and \textit{Photo Editor}~\cite{PhotoEditor}) present data in specialized forms such as map nodes or canvas elements, making it difficult for existing approaches to effectively trigger related DMFs. The remaining case (\textit{CPU Info}~\cite{CPUInfo}) involves a single-object data structure instead of a list-based structure, limiting the applicability of the DUM identification strategy. Future work can further explore a more general and robust paradigm for DUM identification across diverse app types.
Second, although our \textcolor{black}{state-aware} mechanism has achieved some effect, it still cannot fully eliminate hallucinations in UI action plans. As a result, our approach remains limited compared with manually crafted scripts, and 24.3\% DMFs cannot yet be reliably triggered.
For example, in \textit{Home Medkit}~\cite{HomeMedkit}, when generating a Create DMF, a dialog for asking confirmation of ``save medkit group'' is displayed. However, LLMs mistakenly consider that the DMF has been successfully triggered, leading it to prematurely terminate the process. Future work can continue to explore how to enable LLMs to accurately understand the semantics of apps' UI pages, thereby helping to enhance the effectiveness of detecting logical bugs in Android apps.

\begin{mybox}
    \textbf{Answer of RQ1:} LDMDroid achieved an average per-app DMF coverage of 75.7\% and success rate of 62.5\%. It requires only \$0.026 and 153.4 seconds on average to successfully trigger one DMF, demonstrating substantially improved effectiveness and efficiency over the baselines.
\end{mybox}

Table~\ref{tab:dmf-success-rates} also presents the results of an ablation study of LDMDroid, in which the UI change summarization module (LDMDroid w/o UI changes) and the progress tracking module (LDMDroid w/o Tracking Progress) were removed.
In terms of effectiveness, removing either module leads to a noticeable performance drop.
When the UI change summarization module is removed, the per-app average DMF coverage drops to 64.4\%, and the success rate decreases to 33.3\%.
Similarly, removing the progress tracking module results in an average per-app DMF coverage of 54.7\% and success rate of 36.3\%.
We further analyzed several specific cases and found that the removal of either module often leads to UI event sequences deviating from the expected paths to trigger DMFs, leading to fewer DMFs being triggered and lower success rates.

In terms of efficiency, removing either module reduces the overall resource consumption, mainly because complex DMFs fail to be triggered in these variants. As shown in Table~\ref{tab:dmf-success-rates}, LDMDroid without the UI change summarization module requires 27.4K tokens, \$0.017, and 151.3 seconds on average to trigger one DMF.
Removing the progress tracking module results in a consumption of 37.8K tokens, \$0.023, and 126.8 seconds.
Most successfully triggered DMFs in these two variants are relatively simple and involve fewer interaction steps, leading to lower token usage, shorter execution time, and reduced cost.

\begin{mybox}
	\textbf{Answer of RQ2:} The UI change summarization and progress tracking modules enhance LDMDroid’s ability to trigger complex DMFs, resulting in higher DMF coverage and success rates. Although they slightly increase token usage, execution time, and cost, they provide a favorable trade-off between effectiveness and efficiency.
\end{mybox}

\subsection{Results for RQ3} \label{sec:RQ3}

\begin{table}[t]
	\centering
	\caption{Comparison of Bug Reports}
	\label{tab:bug-reports}
	\begin{threeparttable}
		\begin{tabular}{
        p{3cm}|>{\centering\arraybackslash}p{3cm}|>{\centering\arraybackslash}p{2.5cm}|>{\centering\arraybackslash}p{2.5cm}
        }
			\toprule
			\textbf{Approaches}
            & \textbf{TPR (\%)}
            & \textbf{\#Report}
            & \textbf{\#Bug} \\
			\midrule
			\textbf{LDMDroid} & \textbf{61.6 [55.5-67.3]} & \textbf{255} & \textbf{17} \\
			LDMDroid w/o DUM & 39.5 [34.0-45.3] & 281 & 11 \\
			Guardian & 25.6 [19.7-32.4] & 180 & 3 \\
			\textcolor{black}{Genie} & \textcolor{black}{32.6 [20.5-47.5]} & \textcolor{black}{43} &  \textcolor{black}{2} \\
            \textcolor{black}{Odin} & \textcolor{black}{29.2 [20.8-39.4]} & \textcolor{black}{89} &  \textcolor{black}{2} \\
            DMSDroid & \textit{N/A} & 0 &  0 \\
			Fastbot2 & \textit{N/A} & 3,602 &  4 \\
			\bottomrule
		\end{tabular}
		\begin{tablenotes}
			\footnotesize
			\item The ``TPR'' columns show the overall TP rate with 95\% confidence intervals; Fastbot2 and DMSDroid only detect crashes, so TPR does not apply.
		\end{tablenotes}
	\end{threeparttable}
\end{table}

As shown in Table~\ref{tab:bug-reports}, LDMDroid generated a total of 255 bug reports across the three runs on 24 apps, from which we distilled 17 unique bugs, including 6 non-crashing bugs and 11 crash bugs.
\textcolor{black}{
LDMDroid w/o DUM detected 2 non-crashing bugs and 9 crash bugs. The reduced bug detection capability is primarily due to the absence of DUM information, which resulted in scenarios where DMEs were triggered but not properly recognized.
Guardian, Genie, Odin, and Fastbot2 identified only 3, 2, 2, and 4 crash bugs, respectively, while DMSDroid failed to detect any bugs. The limited effectiveness of these baselines mainly stems from their low success rate in triggering DMFs, which restricts coverage of critical execution paths that may expose DMEs. Although Guardian leverages LLM guidance to purposefully trigger DMFs, its success rate remains relatively low (see Table~\ref{tab:dmf-success-rates}). DMSDroid, Genie, Odin, and Fastbot2 rely on randomized event generation, making it difficult to cover the long and specific event sequences required to trigger DMEs.
Furthermore, although Genie and Odin incorporate general oracles for detecting non-crashing bugs, their effectiveness in identifying DMEs remains limited. Genie’s independent view property assumption is not well suited for many DMEs. Odin relies on a behavior clustering based heuristic oracle but ignores GUI textual information, thereby missing DMEs related to text display.
Finally, we found that all bugs discovered by the baselines were also identified by LDMDroid.
}

\begin{figure}[t]
    \centering
    \includegraphics[width=\textwidth]{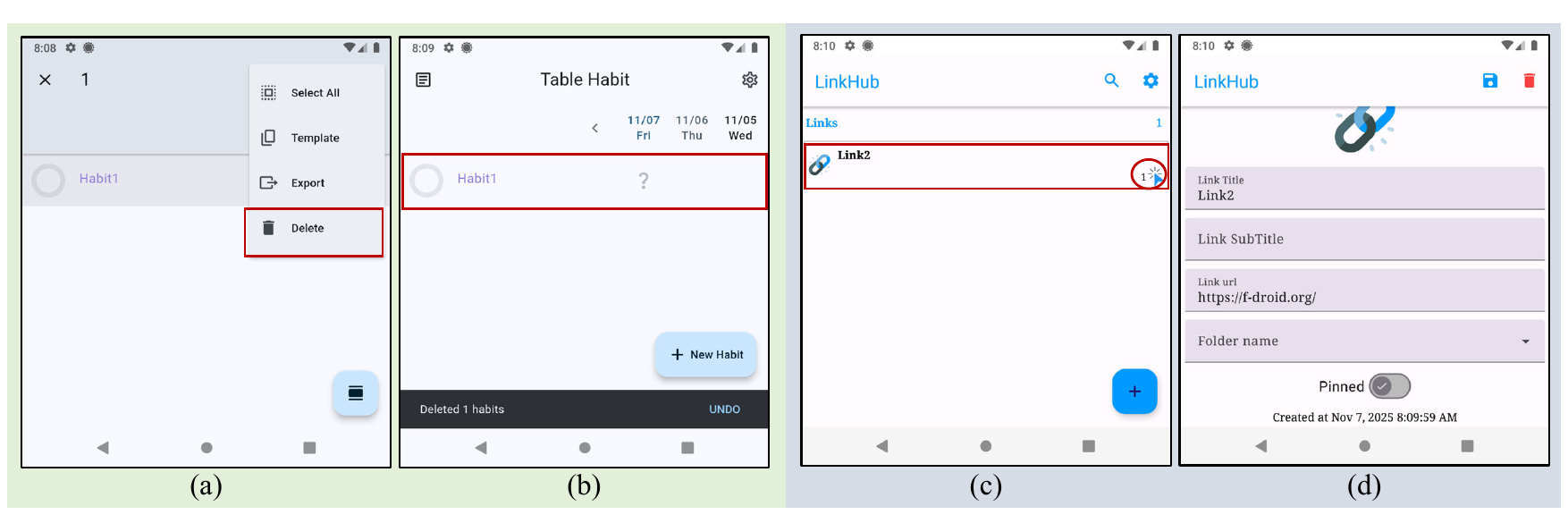}
    \caption{Illustrative examples of LDMDroid detections. (a,b) show a true positive: a deleted habit item remains visible in the list, revealing a logic DME. (c,d) show a false positive: a mismatch between the summary view and detail page was incorrectly flagged, though no DME occurred.}
    \Description{Four screenshots demonstrating LDMDroid detection outcomes. (a) and (b) depict a true positive bug in the Table Habit app, where a deleted habit item remains visible after deletion. (c) and (d) depict a false positive case in the LinkHub app, where a number present in the summary view is absent in the detail page, mistakenly reported as a bug.}
    \label{fig:case-study}
\end{figure}

In terms of precision, both LDMDroid, LDMDroid w/o DUM and Guardian incorporate LLM-based test oracles for DMFs.
However, due to hallucination issues in LLMs, some false positives may occur. Compared to Guardian, LDMDroid mitigates the hallucination by leveraging runtime information related to DUMs to assist the LLM in finding logical DMFs, increasing the TPR from 25.6\% to 61.6\%. Guardian has a lower TPR and detects fewer DMEs due to generating fewer valid DMF instances and inaccurately identifying DUMs, which prevents its test oracle from verifying whether DUM changes are consistent with DMF logic.
LDMDroid w/o DUM removes the DUM identification module, which causes the TPR to drop from 61.6\% to 39.5\%. It also generates more false positive reports due to the absence of DUM information.
Since prior work has not incorporated DUM identification into the automated oracle, this finding highlights the value of our DUM-aware design.
\textcolor{black}{
Although the general oracles used by Genie and Odin are theoretically applicable to a wider range of bugs, their false-positive rate remains high, which limits their effectiveness.
}
Fastbot2 and DMSDroid are limited to detecting crash bugs, which makes false positives unlikely, and their accuracies are therefore not included in the comparison.

Fig.~\ref{fig:case-study}(a) and (b) illustrate a validated DME uncovered by LDMDroid. This DME was found in \textit{Table Habit}~\cite{TableHabit}, an app with over 600 stars that helps users develop habits. Specifically, after LDMDroid completed a delete action on a habit item (see Fig.~\ref{fig:case-study}(a)), the habit still remained visible in the list (see Fig.~\ref{fig:case-study}(b)), resulting in an unexpected behavior manifested in apps' UI.
On the other hand, LDMDroid also generated false bug reports, primarily due to gaps between the UI semantics understood by LLMs and the actual semantics. Fig.~\ref{fig:case-study}(c) and (d) show a case in \textit{LinkHub}~\cite{LinkHub}. After entering the detail page of a specific link to perform a read action, the LLM concluded that there was an inconsistency between the summary shown in the list view (see Fig.~\ref{fig:case-study}(c)) and the information shown on the detail page (see Fig.~\ref{fig:case-study}(d)). Specifically, the number ``1'' displayed in the summary disappeared in the detailed information. However, such inconsistencies did not cause a functional error in the app's UI page.

We submitted bug reports regarding 17 reproducible DMEs detected by LDMDroid. To help app developers fully understand our submitted bug reports, we provide expected behavior and the actual behavior along with detailed reproduction steps and a screen recording video. We \textbf{complied with the app projects' contributing guidelines and licenses} to submit issue reports.

\begin{table}[t]
	\centering
    \small
	\caption{Submitted Bugs of LDMDroid}
	\label{tab:bugs}
	\begin{tabular}{l|l|l|l|l|l}
		\toprule
		\textbf{App Name} & \textbf{ID} & \textbf{State} & \textbf{Related DMFs} & \textbf{Type} & \textbf{Description} \\
		\midrule
		Material Files & \#1392 & Confirmed & Create & Non-crash & File list doesn't refresh after file operations. \\
		Material Files & \#1394 & Pending & Search, Create & Non-crash & Search results do not auto-refresh. \\
		Material Notes & \#416 & Fixed & Create, Read & Non-crash & Incorrect format when sharing note. \\
		PFA Todo List & \#158 & Fixed & Create, Delete & Non-crash & Settings entry mistaken for group item. \\
		Fridgey & \#44 & Fixed & Create, Delete & Crash & Crash from random selection on empty food list. \\
		Fridgey & \#46 & Fixed & Create & Crash & Crash caused by no file selected during import/export. \\
		Fridgey & \#47 & Fixed & Create & Crash & Crash importing an empty data file. \\
		Print Notes & \#12 & Fixed & - & Crash & White screen from missing permission. \\
		To Don't & \#342 & Confirmed & Create & Crash & Crash saving habit without selecting a label. \\
		To Don't & \#343 & Confirmed & Create, Delete & Crash & Crash deleting habit. \\
		To Don't & \#352 & Pending & Create, Update & Crash & Crash setting one-time reminder. \\
		Play NotePad & \#102 & Pending & Create & Crash & Crash adding voice recording. \\
		Play NotePad & \#103 & Pending & Create & Non-crash & Notes remain invisible after label deletion. \\
		NotallyX & \#569 & Fixed & Create, Update & Crash & Crash editing completed item in checklist note. \\
		NotallyX & \#570 & Fixed & Create & Crash & Crash selecting new note while previous is invisible. \\
		Table Habit & \#270 & Pending & Delete & Non-crash & Deleted habit remains visible until list is refreshed. \\
		Read You & \#1019 & Pending & Update & Crash & Crash opening invalid RSS feed link. \\
		\bottomrule
	\end{tabular}
\end{table}

Table~\ref{tab:bugs} summarizes the bugs we submitted.
So far, 14 bugs have been confirmed, among which 11 have been fixed, while the rest are still pending (none have been rejected). These results show that LDMDroid can provide useful information for app developers to detect previously-unknown DMEs in Android apps.

Notably, LDMDroid is capable of generating bug reports whose root causes are located deeply in the UI framework. These issues often require specific environments to be triggered, showing the usefulness of LDMDroid in detecting bugs that are difficult for app developers to identify manually. For example, we submitted an issue report~\cite{IssueReport} to the developer of \textit{PFA Todo List}~\cite{PFATodoList} and received the following response: \noindent \textit{Unfortunately I was not able to fix it but found a workaround. It seems to be a bug in the UI framework. But the user should no longer see the wrong behavior.}

\begin{mybox}
	\textbf{Answer of RQ3:} LDMDroid successfully detects 17 previously-unknown DMEs, achieving a precision of 61.6\%. So far, 14 DMEs have been confirmed and 11 have been fixed by app developers.
\end{mybox}

\subsection{Results for RQ4} \label{sec:RQ4}

\begin{table}[t]
    \centering
    \caption{Comparison of LDMDroid and PBFDroid}
    \label{tab:PBFDroid-compare}
    \begin{threeparttable}
    \begin{tabular}{
    p{2cm}|p{4cm}|>{\centering\arraybackslash}p{3cm}|>{\centering\arraybackslash}p{1cm}
    }
        \toprule
        \textbf{Approach} & \textbf{Manual effort category} & \textbf{Time (s)} & \textbf{\#Bug} \\
        \midrule
LDMDroid & False-positive filtering & 173.9 [155.0-192.9] & 17 \\
\textcolor{black}{PBFDroid} & \textcolor{black}{DMF property specification} & \textcolor{black}{627.7 [515.8-739.5]} & \textcolor{black}{18} \\
        \bottomrule
    \end{tabular}
    \begin{tablenotes}
        \footnotesize
        \item The ``Time'' column reports the per-app average time together with its 95\% confidence intervals.
    \end{tablenotes}
    \end{threeparttable}
\end{table}

\textcolor{black}{
Table~\ref{tab:PBFDroid-compare} presents the comparison results.
In terms of manual effort, LDMDroid required an average of 173.9 seconds per app for false-positive filtering, whereas PBFDroid required an average of 627.7 seconds per app for DMF property specification, indicating a lower manual overhead for LDMDroid.
In terms of bug detection effectiveness, LDMDroid detected 17 bugs and PBFDroid detected 18 bugs, with 16 bugs overlapping. LDMDroid additionally identified one non-crashing bug (\textit{PFA Todo List \#158}), which contributes to the LLM's generalization capability, suggesting LDMDroid is able to uncover DMEs beyond predefined DMF configurations. PBFDroid detected one additional crashing bug (\textit{CycleStreets \#565}) and one additional non-crashing bug (\textit{Read You \#1073}), both associated with DMFs not triggered by LDMDroid.
Overall, the results indicate that LDMDroid is practical in terms of manual effort and bug detection capability.
}

\begin{mybox}
	\textbf{Answer of RQ4:} \textcolor{black}{LDMDroid requires less manual effort while maintaining comparable bug detection capability. Participants spent on average 173.9 seconds per app with LDMDroid, compared to 627.7 seconds per app with PBFDroid. LDMDroid detected 17 bugs and PBFDroid detected 18 bugs, with 16 bugs overlapping.}
\end{mybox}

\subsection{Model Sensitivity Analysis}

\textcolor{black}{
As shown in Table~\ref{tab:llm-comparison}, LDMDroid maintains broadly consistent performance across different multimodal models, suggesting that the results are not substantially affected by model selection.
}

\begin{table}[t]
    \centering
    \caption{\textcolor{black}{Model Sensitivity Analysis of LDMDroid's Effectiveness}}
    \label{tab:llm-comparison}
    \begin{threeparttable}
    \begin{tabular}{
    p{2cm}|>{\centering\arraybackslash}p{3cm}|>{\centering\arraybackslash}p{3cm}|>{\centering\arraybackslash}p{3cm}
    }
        \toprule
        \textbf{Model} & \textbf{$\bm{\mathrm{DMF}_{\mathrm{cov}}}$ (\%)} & \textbf{SR (\%)} & \textbf{TPR (\%)}\\
        \midrule
GLM-4V-Plus & 75.7 [62.7-88.7] & 62.5 [57.0-68.0] & 61.6 [55.5-67.3] \\
GPT-4o & 78.6 [69.8-87.5] & 67.3 [59.2-75.4] & 70.1 [54.8-85.4] \\
Qwen2.5-VL & 69.3 [55.6-83.0] & 58.7 [44.7-72.7] & 55.9 [40.6-71.2] \\
        \bottomrule
    \end{tabular}
    \end{threeparttable}
\end{table}

\section{Threats to Validity}
The primary threat to external validity lies in the limited representativeness of the evaluation subjects. To mitigate potential bias in app selection, we ensured diversity by choosing apps with a range of GitHub star distributions. To avoid potential overfitting, we excluded subjects that were already included in the PBFDroid~\cite{sunPropertyBasedFuzzingFinding2023} dataset, as the domain knowledge of LDMDroid was derived from them. Additionally, we selected apps that have been recently updated to ensure our evaluation results are not outdated and to increase the likelihood of receiving feedback from app developers.

The main threat to internal validity arises from the manual effort involved in the evaluation, which may introduce subjectivity or human error. To mitigate this risk, we adopted the following measures to enhance the accuracy and consistency of manual validation:
(1) For the tasks involving human judgment of correctness, two authors independently conducted the evaluation, with disagreements resolved by a third author.
For the assessment of manual effort, we provided training to the recruited participants, and each app was evaluated independently by five different participants.
(2) We submitted only those bugs that could be consistently reproduced across multiple runs, thereby avoiding submissions to app developers that cannot be reproduced and could otherwise clutter the open-source community.
\section{Related Work}

To ensure the quality of Android apps, many researchers have been investigating automated app testing techniques~\cite{kongAutomatedTestingAndroid2019, androidMonkey2025, liDroidBotLightweightUIguided2017, huang2018understanding, huang2021characterizing, huang2023conffix, gu2025characterizing, hu2018tale,suGuidedStochasticModelbased2017a, guPracticalGUITesting2019, wangComboDroidGeneratingHighquality2020a, liHumanoidDeepLearningbased2020, panReinforcementLearningBased2020a, lvFastbot2ReusableAutomated2023}.
Specifically, a set of tools mainly rely on random or rule-based approaches. For example, Monkey~\cite{androidMonkey2025} is a popular random-based automated Android UI testing tool that performs testing by generating random UI and system events; 
DroidBot~\cite{liDroidBotLightweightUIguided2017} adopts a rule-guided random strategy to explore the UI and generate a UI transition graph (UTG).
There are also a set of model-based automated testing approaches~\cite{suGuidedStochasticModelbased2017a, guPracticalGUITesting2019, wangComboDroidGeneratingHighquality2020a, liHumanoidDeepLearningbased2020, panReinforcementLearningBased2020a, lvFastbot2ReusableAutomated2023},
which analyze the UI transitions of apps and adopt different strategies to guide the testing process.
For instance, Fastbot2~\cite{lvFastbot2ReusableAutomated2023} leverages its model to maintain a memory of previously covered activity paths during testing, which guides subsequent exploration.
While the above tools mainly focus on detecting crash bugs, there also exist approaches designed to automatically detect non-crashing bugs in Android apps.
For example, Genie~\cite{Genie} introduces an independent view–based functional fuzzing approach that leverages the independence among sibling views in Android GUIs to automatically generate property-preserving mutant tests.
Odin~\cite{Odin} adapts the ``bugs as deviant behaviors'' paradigm to Android testing, using unsupervised deep-state differential analysis to automatically detect non-crashing functional bugs.
Although these automated Android testing tools have improved testing efficiency, their generated test behaviors are still driven by random or heuristic rules. As a result, they lack a true understanding of UI semantics and cannot reason about interactions that reflect actual user intentions.

LLMs have achieved remarkable progress in natural language processing~\cite{zhaoSurveyLargeLanguage2025}, and some studies have begun applying LLMs to automated Android UI testing.
For example, AutoDroid~\cite{wenAutoDroidLLMpoweredTask2024} constructs a UTG through pre-exploration of the app and uses LLMs to summarize the functions of different UI elements, thereby assisting LLMs in completing mobile tasks.
Guardian~\cite{ranGuardianRuntimeFramework2024} refines the set of available UI actions and incorporates a reflection strategy to enhance the LLMs' UI exploration capabilities.
VisionTasker~\cite{songVisionTaskerMobileTask2024} combines image understanding with LLMs to support mobile task completion.
Mobile-Agent-E~\cite{wangMobileAgentESelfEvolvingMobile2025} designs a multi-agent hierarchical collaboration mechanism, where multiple LLM agents work together to accomplish complex mobile tasks. InputBlaster~\cite{liuTestingLimitsUnusual2024} generates unusual text inputs using LLMs to test the robustness of apps.
GPTDroid~\cite{liuMakeLLMTesting2024} explores various functionalities of an app through LLMs to enable comprehensive testing.

However, existing approaches face two critical limitations when it comes to detecting DMEs:
First, they lack sufficient knowledge on (1) the triggering of DMFs, and (2) the test oracles for identifying DMEs. To address the above limitations, LDMDroid introduces \textcolor{black}{state-aware} reasoning to monitor the runtime behavior of DMFs, and leverages such runtime behavior as the test oracle for identifying DMEs, therefore improving the performance of LLMs in DME detection tasks.

\noindent \textbf{Detecting DMEs in Android Apps.}
Our work focuses on detecting DMEs in Android apps. Specifically, PBFDroid~\cite{sunPropertyBasedFuzzingFinding2023} and Kea~\cite{xiongGeneralPracticalPropertybased2024} are both semi-automated approaches to detect DMEs in Android apps. However, both approaches rely on manual effort to build test scripts that can trigger DMFs for detecting DMEs.
Gu et al. introduced DMSDroid~\cite{guDMSDroid2024}, a tool designed to automatically collect DMF instances to aid in DME detection. Specifically, DMSDroid utilizes DroidBot~\cite{liDroidBotLightweightUIguided2017} to pre-explore the app and build a UTG, employing hardcoded rules to identify potential DMFs. However, the random event generation strategy of DroidBot results in limited UTG coverage, and the hardcoded rules in DMSDroid only encompass a small subset of DMFs. Consequently, these limitations prevent it from being effectively generalized across various Android apps. In view of the above limitations, we propose LDMDroid, which leverages the semantic reasoning capabilities of LLMs to enable fully automated DME detection. Experimental results demonstrate that LDMDroid significantly outperforms existing tools in both DMF collection and bug detection (see Section~\ref{sec:evaluation}).
\section{Conclusion}

In this paper,  we propose an LLM-based approach called LDMDroid for automatic detection of DMEs in Android apps. 
Specifically, LDMDroid utilizes a \textcolor{black}{state-aware} mechanism to guide LLMs through the step-by-step reasoning process for generating UI event sequences, which improves the success rates of triggering DMFs. LDMDroid then employs general visual features to automatically detect changes in the states of the data being manipulated, thereby enhancing the accuracy of LLMs in verifying DMEs.
We implemented the proposed approach in a tool named LDMDroid and evaluated it on 24 real-world open-source Android apps.
LDMDroid demonstrates substantially improved effectiveness and efficiency compared with existing approaches, highlighting its strong capability in triggering DMFs.
Furthermore, LDMDroid discovered 17 unique bugs across 10 apps. As of now, 14 bugs have been confirmed by developers, and 11 have been fixed.
For future work, we plan to extend LDMDroid to identify and handle more types of logic errors in Android apps, such as issues related to numerical computations, and temporal logic.

\section{Data Availability}

We release the implementation of LDMDroid, together with the discovered DMF instances and reported bugs, on the project website \href{https://github.com/runnnnnner200/LDMDroid}{https://github.com/runnnnnner200/LDMDroid}.

\section*{Acknowledgment}

This work was supported by the National Natural Science Foundation of China (Grant No. 62402405), Fujian Provincial Natural Science Foundation of China (Grant No. 2026J001003), Xiamen Natural Science Foundation (Grant No. 3502Z202471016), and the Fundamental Research Funds for the Central Universities (Grant No. 20720240087, 20720250029). Huaxun Huang is the corresponding author and works as a member of Xiamen Key Laboratory of Intelligent Storage and Computing in Xiamen University.

\bibliographystyle{ACM-Reference-Format}
\bibliography{reference}

\end{document}